\documentclass[pre,twocolumn]{revtex4}
\usepackage{epsfig}
\usepackage{bm}
\usepackage{amssymb}
\usepackage{amsmath}

\begin{document}

\title{From turbulence to deterministic chaos in freely decaying fluid dynamics}

\author{A. Bershadskii}

\affiliation{
ICAR, P.O. Box 31155, Jerusalem 91000, Israel
}

\begin{abstract}

 The transition from hard/soft turbulence to deterministic chaos in freely decaying fluid dynamics (incompressible and compressible) has been studied using the results of laboratory measurements and numerical simulations. The notion of distributed chaos has been applied in order to quantify differences in the intermediate regimes appearing during free decay when hard/soft turbulence is eventually decaying to the state of deterministic chaos.  Free decay in magnetohydrodynamics has been also briefly discussed in this context (with an application to the results of measurements in the solar photosphere).  
\end{abstract}

\maketitle

\section{Introduction}

  Free decay (without external forcing) is usually considered as one of the simplest chaotic/turbulent flows. However, the numerous laboratory experiments and numerical simulations show a substantial variability of the free decay properties. I will be shown in the present study that this variability is mainly based on the spontaneous breaking of local reflectional symmetry and the abundance of the invariants (ideal and dissipative) in the free decay. The differences in the initial/boundary conditions and in values of Reynolds number turned out to be significant factors controlling these phenomena. \\
  
  Historically, most of the previous studies of free decay were concentrated on the decay of the mean kinetic energy with time, but no universal decay laws were recognized (see for excellent recent reviews Refs. \cite{ym},\cite{jds}). Maybe it is related to the natural relation of free decay of the mean kinetic energy to the large-scale dynamics, which is crucially dependent on the initial/boundary conditions and the finite Reynolds numbers (the latter are generally changing with the time of the decay). The intermediate- and small-scale dynamics in free decay were much less under consideration in those studies. However, free decay can be used for studying the fundamental fluid dynamics just on these scales due to the absence of complications related to the external forcing.\\
 
 For instance, for buoyancy-driven (i.e. with an external forcing) convection three main dynamical regimes for these scales were recognized according to the experimental observations \cite{hcl}. Namely, they are deterministic chaos, `soft' turbulence, and `hard' turbulence.  The `hard' turbulence was related to the existence of an inertial range of scales with a scaling (power) spectral law, whereas the `soft' turbulence was vaguely defined.  It is shown in the present paper that these three main regimes can be recognized and studied also in free decay.\\

\begin{figure} \vspace{-0.3cm}\centering 
\epsfig{width=.45\textwidth,file=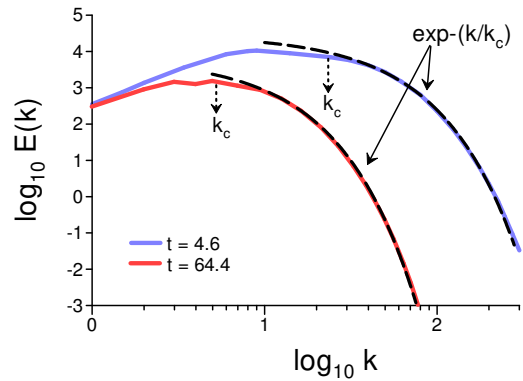} \vspace{-4.5cm}
\caption{Energy spectra computed in the free isotropic decay at $t = 4.46$ and $t = 64.4$ (the final time of the DNS) at small $Re_{\lambda}$.} 
\end{figure}

  From a dynamic point of view, these regimes can be based on the notion of smoothness. Namely, the non-smooth dynamics is associated with the scaling (power-law) spectra, whereas smooth dynamics is usually associated with the {\it stretched} exponential spectra. In particular, the deterministic chaos is typically associated with smooth trajectories (which are sensitive to the initial conditions) and is characterized by {\it exponential} power spectra \cite{fm}-\cite{kds}. Obviously, there can exist chaotic-like dynamics which is different from the deterministic chaos but still smooth and characterized by {\it stretched} exponential spectrum. This type of chaotic-like dynamics (containing elements of randomness) is a good candidate for the role of the `soft' turbulence separating between the deterministic chaos and the `hard' turbulence. This dynamics (which is based on the notion of distributed chaos) will be used in the present paper for studying the intermediate regimes of free decay. Distributed chaos is much more reach than deterministic chaos and characterized by different intermediate regimes (which are dominated by different dynamical invariants).

 \section{Deterministic chaos in free decay}  

   It was already mentioned in the Introduction that for the deterministic chaos (with smooth trajectories) the power spectra typically have exponential form \cite{fm}-\cite{kds} 
$$ 
E(k) \propto \exp(-k/k_c)  \eqno{(1)}
$$
 where $k_c$ can be considered as a characteristic wavenumber for the system. \\
 
   Since in fluid dynamics deterministic chaos can be expected at sufficiently small Reynolds numbers, the later stages of the free decay are the most suitable situations for studying the transition from random to deterministic (chaotic) dynamics. \\
   
   Let us begin with the free decay characterized by {\it small} Reynolds numbers already at an earlier stage. Results of a DNS for the free decay with small Reynolds numbers ($Re_\lambda < 20$, where $Re_{\lambda}$ is the Taylor-Reynolds number) were reported in a recent paper Ref. \cite{ajv} (cf Ref. \cite{kds} for the externally forced case). 
   
     The Navier–Stokes equation
 $$
 \frac{\partial {\bf u}}{\partial t} = - ({\bf u} \cdot \nabla) {\bf u} 
    -\frac{1}{\rho} \nabla {p}  + \nu \nabla^2  {\bf u} \eqno{(2)}
$$
for an incompressible fluid $ \nabla \cdot {\bf u} = 0$ was numerically solved with  the initial conditions taken as a random Gaussian noise with an energy spectrum in the standard form 
$$
E(k) \propto k^n  \exp(-k/k_0)^2  \eqno{(3)}
$$
where $n = 2$. The Eq. (2) was solved in a spatially periodic cubic box \cite{ajv}. \\

  Figure 1 shows the energy spectra computed at $t = 4.46$ and $t = 64.4$ (the final time of the DNS). The spectral data were taken from Fig. 5a of the Ref. \cite{ajv}. The dashed curves indicate the exponential spectrum Eq. (1) and the dotted arrow indicates the position of the characteristic scale $k_c$.\\
 
\begin{figure} \vspace{-0.48cm}\centering
\epsfig{width=.44\textwidth,file=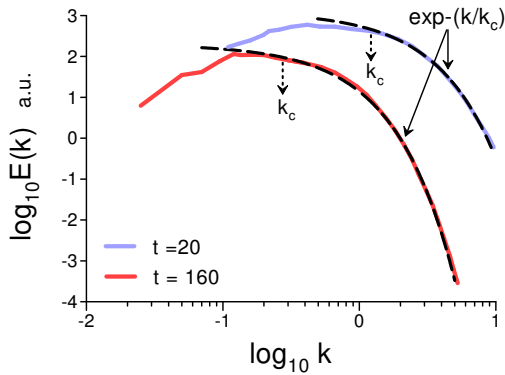} \vspace{-4.54cm}
\caption{Energy spectra computed in the free helical decay at $t = 20$ and $t = 160$ (the final time of the DNS) at small Reynolds numbers.} 
\end{figure}
\begin{figure} \vspace{-0.61cm}\centering\hspace{-1.8cm}
\epsfig{width=.45\textwidth,file=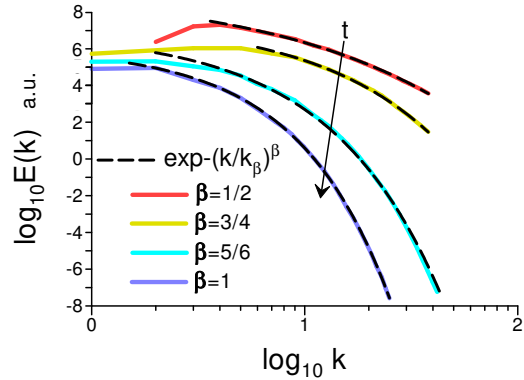} \vspace{-4.65cm}
\caption{ The energy spectrum decay for a DNS with initial (at $t =0$) $Re_{\lambda} \simeq 72.37$. }
\end{figure}

   The previous example was isotropic, homogeneous, and nonhelical (with global reflectional symmetry). In a recent paper Ref. \cite{yk} results of an analogous DNS with the small Reynolds numbers were reported, but in this case the dynamics was helical from the very beginning.   Figure 2 shows the energy spectra computed at $t = 20$ and $t = 160$ (the final time of the DNS). The spectral data were taken from Fig. 7a of the Ref. \cite{yk}. The dashed curves indicate the exponential spectrum Eq. (1) and the dotted arrow indicates the position of the characteristic scale $k_c$.\\
    
     Results of a pseudo-spectral DNS of the free isotropic decay in a periodic cube with initial Taylor-Reynolds number $Re_{\lambda} \simeq 72.37$  were reported in a paper \cite{peng}. The energy spectrum of the initial random noise used for this DNS was taken in the standard form Eq. (3) with $n = 4$ and was concentrated in a large-scale range of scales. \\  
     
     Figure 3 shows a picture of the energy spectrum decay for the DNS. The spectral data were taken from Fig. 22 of the Ref. \cite{peng}. The lowest dashed curve indicates the exponential spectrum (deterministic chaos) and the upper dashed curves indicate stretched exponential spectra. The stretched exponential spectra will be discussed below. \\
     
   Results of a DNS of the free isotropic decay in a periodic cube with a larger initial Taylor-Reynolds number $Re_{\lambda} =143$ were reported in a recent paper \cite{ym}. The energy spectrum of the initial random noise used for this DNS was again taken in the standard form Eq. (3) with $n = 4$. \\ 
   
   Figure 4 shows a picture of the energy spectrum decay for the DNS. The spectral data were taken from Fig. 14 of the Ref. \cite{ym}. The $k_p$ corresponds to the peak value of the initial energy spectrum. The decreasing $Re_{\lambda}$ corresponds to the increasing time ($t$) of the free decay. The dashed curves indicate the exponential spectrum (deterministic chaos for the small $Re_{\lambda} =18$, cf Ref. \cite{kds}) and stretched exponential spectra for larger $Re_{\lambda}$ (smaller $t$). \\
   
   In recent papers Refs. \cite{znw},\cite{kita} results of laboratory experiments behind grids in a wind tunnel with small and moderate $Re_{\lambda}$ were reported. It is commonly believed that such experiments can mimic free decay. The $Re_{\lambda}$ was decreasing with the distance from the grid $x$, and in the `free decay' interpretation $x \rightarrow t$. \\
   
     Figure 5 shows the power spectra of the measured streamwise velocity fluctuations vs streamwise wavenumber for small and moderate $Re_{\lambda}$ (the spectra are shifted for clarity). The spectral data were taken from Fig. 6 of the Ref. \cite{znw}. The dashed curves indicate the exponential spectrum (deterministic chaos) for the small $Re_{\lambda} =5$ and stretched exponential spectra for moderate $Re_{\lambda}$ (cf Fig. 4). It should be noted that for a small (see above) $Re_{\lambda} = 14$ the spectrum is still stretched exponential. It means that the criterion of `smallness' of $Re_{\lambda}$ can also depend on the initial/boundary conditions. We will return to the discussion of these experiments below.

\begin{figure} \vspace{-0.9cm}\centering\hspace{-1.8cm}
\epsfig{width=.45\textwidth,file=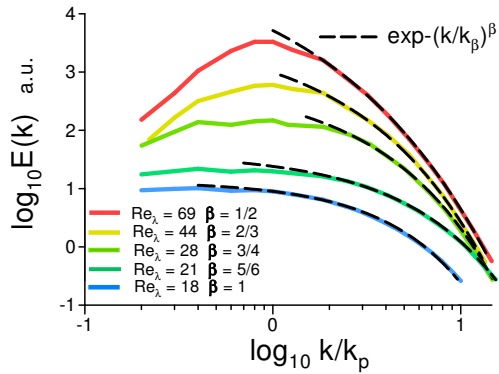} \vspace{-4.23cm}
\caption{ The energy spectrum decay for a DNS with moderate and small $Re_{\lambda}$ (the spectra are shifted for clarity). The initial value of $Re_{\lambda} = 143$. }
\end{figure}
\begin{figure} \vspace{-0.22cm}\centering \hspace{-1cm}
\epsfig{width=.45\textwidth,file=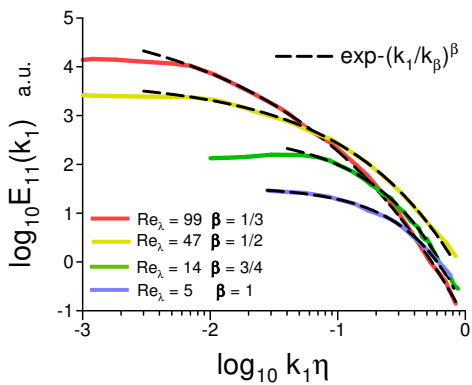} \vspace{-4.62cm}
\caption{The power spectra of the measured behind a grid streamwise velocity fluctuations vs streamwise wavenumber (the spectra are shifted for clarity).}
\end{figure}

\section{Spontaneous breaking of local reflectional symmetry}   
   
    The global reflection symmetry results in zero mean/global helicity. However, the point-wise helicity can be not identically equal to zero in this case. Spontaneous breaking of the local reflectional symmetry (and related to this phenomenon spontaneous helicity fluctuations) is a generic property of the chaotic/turbulent flows (see for instance Ref. \cite{kerr},\cite{hk}). The appearance of the moving with the fluid vorticity blobs having non-zero blob's helicity can accompany this process \cite{moff1}-\cite{bt}, and eventually, all the localized patches of the nonzero helicity can be represented by the vorticity blobs. Since the global helicity should be still equal to zero these localized positive and negative blob's helicities have to be canceled at the overall average. At the boundaries of the vorticity blobs the vorticity  field ${\boldsymbol \omega}$ is tangential (i.e. ${\boldsymbol \omega} \cdot {\bf n}=0$, where ${\bf n}$ is a unit normal to the boundary) .\\
    
 \subsection{Moments of helicity distribution}

  The helicity in a vorticity blob with spatial volume $V_j$ is
$$
H_j = \int_{V_j} h({\bf r},t) ~ d{\bf r}.  \eqno{(4)}
$$
where $h({\bf r},t) = {\bf v} \cdot {\boldsymbol \omega}$ is helicity distribution, $\bf{u}$ is velocity field, and ${\boldsymbol \omega} ({\bf r},t)= \nabla \times {\bf u}  ({\bf r},t)$ is vorticity field.\\

   The moments of the helicity distribution can be defined as \cite{lt} ,\cite{mt}
$$
{\rm I_n} = \lim_{V \rightarrow  \infty} \frac{1}{V} \sum_j H_{j}^n  \eqno{(5)}
$$
here $V$ denotes the total volume of the vorticity blobs.\\

Due to the global reflectional symmetry all odd moments are identically equal to zero. \\

   Let us denote the helicity of the blobs having negative helicity as $H_j^{-}$, and the helicity of the blobs having positive helicity as $H_j^{+}$, and let us denote corresponding moments as
$$
{\rm I_n^{\pm}} = \lim_{V \rightarrow  \infty} \frac{1}{V} \sum_j [H_{j}^{\pm}]^n  \eqno{(6)}
$$ 
where the summation in Eq. (6) is over the blobs with negative (or positive) helicity only.  \\

 The odd moments  ${\rm I_n} = {\rm I_n^{+}} + {\rm I_n^{-}} =0$ (due to the global reflectional symmetry), then ${\rm I_n^{+}} = - {\rm I_n^{-}}$ for odd $n$.\\
 
 All moments of the helicity distribution ${\rm I_n}$ are ideal (nondissipative) invariants both for incompressible and compressible fluids \cite{lt},\cite{mt} (the same is also true for the moments ${\rm I_n^{\pm}}$). 
 
 \subsection{Chkhetiani invariants and their moments}
 
  In ideal (non-dissipative) fluid dynamics the energy and helicity are fundamental invariants. The dissipative Navier-Stokes equations have their own (dissipative) fundamental invariants related to the conservation of linear momentum: Birkhoff-Saffman invariant \cite{bir},\cite{saf},\cite{dav}
$$   
\mathcal{S} = \int  \langle {\bf u} ({\bf x},t) \cdot  {\bf u} ({\bf x} + {\bf r},t) \rangle d{\bf r},  \eqno{(7)}
$$   
and of angular momentum: Loitsyanskii invariant \cite{dav} ,\cite{my}
$$
\mathcal{L} =  \int r^2 \langle {\bf u} ({\bf x},t) \cdot  {\bf u} ({\bf x} + {\bf r},t) \rangle d{\bf r}  \eqno{(8)}
$$  
where $<...>$ denotes a global average.\\

In paper Ref. \cite{otto1} a new class of the invariants (Chkhetiani invariants) of the dissipative Navier-Stokes equations has been added:
 $$
^p\mathcal{I} = \int r^p\langle {\bf u} ({\bf x},t) \cdot  {\boldsymbol \omega} ({\bf x} + {\bf r}, t) \rangle d{\bf r}   \eqno{(9)}
$$  
here  $p =0,1,2$. \\
 
  Let us consider the spontaneous breaking of the local reflectional symmetry and denote 
$$
^p\mathcal{H}_j = \int_{V_j}  r^p\langle {\bf u} ({\bf x},t) \cdot  {\boldsymbol \omega} ({\bf x} + {\bf r}, t) \rangle d{\bf r}  \eqno{(10)}
$$
for the vorticity blobs. Let us also denote $^p\mathcal{H}_{j}^{+}$ and $^p\mathcal{H}_{j}^{-}$ as it was made above for helicity. 
   Then 
 $$
^p\mathcal{I}^{\pm} =   \sum_j~  ^p\mathcal{H}_j^{\pm}    \eqno{(11)}
$$
 where the summation in Eq. (11) is over the blobs with positive (or negative) $^p\mathcal H_j$ only. \\
 
 In the paper Ref. \cite{otto1} the Chkhetiani invariants were proven to be conserved for the isotropic and homogeneous dissipative case. If they are also conserved over the vorticity blobs one can consider $^p\mathcal{I}^{\pm}$ as dissipative invariants as well.\\
 
  We can also introduce the moments based on the $^0\mathcal{H}_j$ (see Eq. (10)) instead of those based on  $H_j$ (see Eq. (4))
$$
{^0\tilde{\mathcal{I}}_n^{\pm}} = \lim_{V \rightarrow  \infty} \frac{1}{V} \sum_j [^0\mathcal{H}_{j}^{\pm}]^n,  \eqno{(12)}
$$
 
 If $^0\mathcal{H}_{j}^{\pm}$ are conserved over the vorticity blobs then the moments ${^0\tilde{\mathcal{I}}_n^{\pm}}$ defined by the Eq. (12) are dissipative invariants.

\section{Introduction to distributed chaos}

 If the parameter $k_c$ in the exponential spectrum Eq. (1) randomly fluctuates, then an ensemble averaging should be used in order to obtain an average spectrum 
$$
E(k) \propto \int_0^{\infty} P(k_c) \exp -(k/k_c)dk_c \eqno{(13)}
$$    
here $P(k_c)$ is a probability distribution of the randomly fluctuating parameter $k_c$. The chaos corresponding to Eq. (13) is not deterministic but it can be still smooth and the spectrum will be stretched exponential. The name `distributed chaos' seems to be an appropriate one for this type of chaos. \\ 

One can use invariants of the flow to find the probability distribution $P(k_c)$. Since the decaying distributed chaos in fluid dynamics has a multitude of invariants one can expect a considerable variability of the corresponding instantaneous $P(k_c)$ during the decay. \\

  Let us consider, as an example, the third moment of the helical distribution ${\rm I}_3^{\pm}$. This case was chosen due to its simplicity (cf. below).

   It follows from the dimensional considerations that
$$
 v_c \propto |I_3^{\pm}|^{1/6}~ k_c^{1/2}    \eqno{(14)}
$$       
 where $v_c$ is a characteristic velocity.
  
   Using Eq. (14) and the normal distribution of $v_c$  \cite{my} one can obtain $P(k_c)$  
$$
P(k_c) \propto k_c^{-1/2} \exp-(k_c/4k_{\beta})  \eqno{(15)}
$$
here $k_{\beta}$ is a constant parameter.\\
 
\begin{figure} \vspace{-0.94cm}\centering \hspace{-1cm}
\epsfig{width=.45\textwidth,file=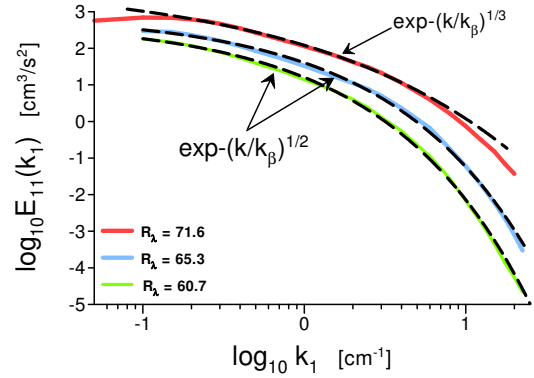} \vspace{-4.45cm}
\caption{The power spectra of the streamwise velocity fluctuations (measured behind a grid) vs streamwise wavenumber for the moderate $R_{\lambda} = 71.6,~65.3,~60.7$.} 
\end{figure}
\begin{figure} \vspace{-0.5cm}\centering  \hspace{-1cm}
\epsfig{width=.5\textwidth,file=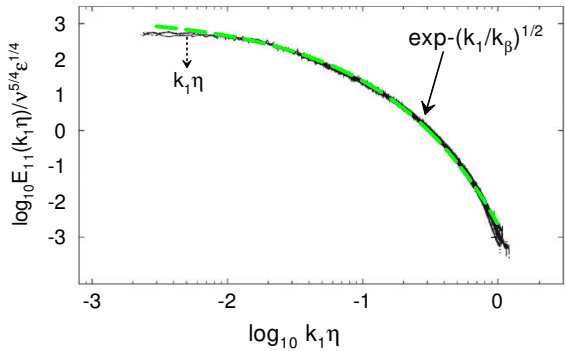} \vspace{-5.9cm}
\caption{ The same as in Fig. 6 but for a composite grid \cite{kda} and for $R_{\lambda} \simeq 67.3$ (from $x/M_L=15$ to $x/M_L =65$). The Kolmogorov variables $\nu$ and $\varepsilon$ have been used for the normalization. } 
\end{figure}
\begin{figure} \vspace{-0.5cm}\centering
\epsfig{width=.45\textwidth,file=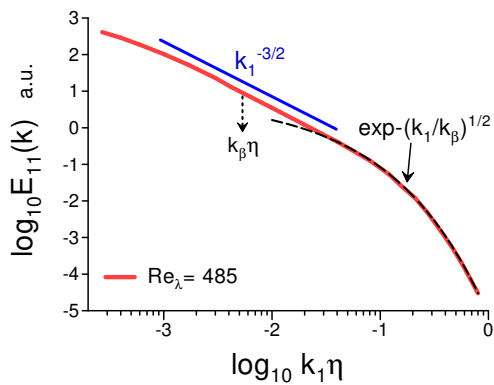} \vspace{-4.35cm}
\caption{ The same as in Fig. 6 but for an active grid \cite{hl} and for $R_{\lambda} = 485$. } 
\end{figure}

    Substituting the probability distribution $P(k_c)$ from Eq. (15) into Eq. (13) one obtains a stretched exponential spectrum
$$
E(k) \propto \exp-(k/k_{\beta})^{1/2}.  \eqno{(16)}
$$     
 
  This result can be extended on a dissipation range of scales by replacing the third moment of helicity distribution $I_3^{\pm}$ with the third moment of the Chkhetiani invariant ${^0\tilde{\mathcal{I}}_3^{\pm}}$ Eq. (12). These invariants have the same dimensionality. Analogous replacements are also valid for other invariants of helicity distribution (see next Section). \\

     In Fig. 4 we have already seen an example of the stretched exponential spectrum Eq. (16) (distributed chaos) observed in a direct numerical simulation \cite{ym} at moderate $R_{\lambda} =69$ (with further decay the spectrum eventually became exponential at a small $R_{\lambda} = 18$ that corresponds to deterministic chaos).\\
     
     An analogous picture is shown in Fig. 5 (for $Re_{\lambda} = 47)$, where one can see the spectra measured in the decaying flow behind a grid \cite{znw} . The wavenumber has been normalized by Kolmogorov's scale $\eta$ and one can see that the fit by the Eq. (16) reaches the near-dissipation range of scales. Therefore, one should consider the third moment of the Chkhetiani invariant ${^0\tilde{\mathcal{I}}_3^{\pm}}$ Eq. (12) (instead of the third moment of the helicity distribution) as a governing invariant in the Eq. (14).\\
     
     Figure 6 shows the power spectra of the streamwise velocity fluctuations (measured behind a grid) vs streamwise wavenumber. The spectra were reported in a classic paper Ref. \cite{cc} for the moderate $Re_{\lambda} = 71.6,~65.3,~60.7$ (Table 2a). The dashed curves indicate correspondence to the stretched exponential spectra (distributed chaos). In particular, for the $Re_{\lambda} = 65.3,~60.7$ the spectra correspond to the Eq. (16). \\
     
  Results of an interesting recent experiment, with free decay behind a grid made up of two closely located perforated plates with the same solidity but different
mesh sizes, were reported in a recent paper Ref. \cite{kda}. Despite decaying kinetic energy the $Re_{\lambda} \simeq 63.7$  along the streamline from $x/M_L=15$ to $x/M_L =65$ (where $x$ is the downflow distance to the grid and $M_L$ is the large mesh size). With the approximately constant $R_{\lambda}$ one can expect a constant value of $\beta$ in the stretched exponential spectrum
$$
E(k) \propto \exp-(k/k_{\beta})^{\beta}  \eqno{(17)}
$$
   Figure 7 shows the power spectra of the streamwise velocity fluctuations measured at different distances from the grid: from $x/M_L=15$ to $x/M_L =65$. The spectral data were taken from Fig. 13b of the Ref. \cite{kda}. The spectra are well collapsed when the normalization by the Kolmogorov variables $\nu$ and $\varepsilon$ have been used. The dashed curve indicates correspondence to the stretched exponential spectrum Eq. (16) (see also comments to the Fig. 5). \\
   
   Figure 8 shows the power spectra of the streamwise velocity fluctuations measured downflow ($x/M = 41$) of an active grid (with rotating wings) at $R_{\lambda} \simeq 485$. The spectral data were taken from Fig. 12 of Ref. \cite{hl}. One can see that at this comparatively large $R_{\lambda}$ a scaling region of scales (corresponding to the nonsmooth dynamics) already appears. On the other hand, the near-dissipation range of scales is characterized by a stretched exponential spectrum Eq. (16) (indicated by the dashed curve), which corresponds to the distributed chaos dominated by the third moment of the Chkhetiani invariant.

\section{Variability of distributed chaos in free decay}    
  
 In general, the estimation Eq. (14) for the spontaneous breaking of local  reflectional symmetry can be replaced by
$$
 v_c \propto  |I_n^{\pm}|^{1/2n}~ k_c^{\alpha_n}   \eqno{(18)}
 $$  
 for the odd moments and by estimation
 $$
  v_c \propto  I_n^{1/2n}~ k_c^{\alpha_n},    \eqno{(19)}
 $$  
 for the even $n$,  where
$$
\alpha_n = 1-\frac{3}{2n},  \eqno{(20)}
$$  
and the stretched exponential  spectrum Eq. (16) can be generalized for the case of smooth dynamics
$$
E(k) \propto \int_0^{\infty} P(k_c) \exp -(k/k_c)dk_c \propto \exp-(k/k_{\beta})^{\beta} \eqno{(21)}
$$  
 
  The distribution $P(k_c)$ can be estimated from Eq. (21) for large $k_c$ \cite{jon}
$$
P(k_c) \propto k_c^{-1 + \beta/[2(1-\beta)]}~\exp(-\gamma k_c^{\beta/(1-\beta)}), \eqno{(22)}
$$     
$\gamma$ is a constant.\\

    A relationship between the exponents $\beta_n$ and $\alpha_n$ can be readily obtained if $v_c$ has Gaussian distribution \cite{my}. In this case it follows from the Eqs. (18,19) and (22)
$$
\beta_n = \frac{2\alpha_n}{1+2\alpha_n}  \eqno{(23)}
$$
 
  Substituting Eq. (20) into Eq. (23) we obtain
 $$
 \beta_n = \frac{2n-3}{3n-3}   \eqno{(24)}  
 $$

  Then for $n \gg 1$
$$
E(k) \propto \exp-(k/k_{\beta})^{2/3},  \eqno{(25)}
$$ 
whereas for $n=2$ (i.e. for the Levich-Tsinober invariant \cite{lt})
$$
E(k) \propto \exp-(k/k_{\beta})^{1/3}  \eqno{(26)}
$$ 
   
   Analogous consideration gives the same results for the moments of the Chkhetiani invariant, i.e. extends these results on a dissipative range of scales. \\
 
\begin{figure} \vspace{-1.3cm}\centering \hspace{-1cm}
\epsfig{width=.47\textwidth,file=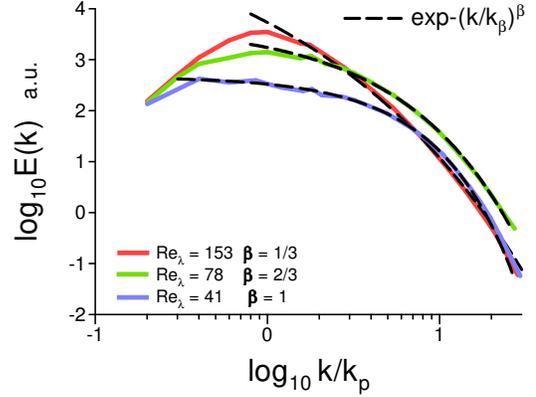} \vspace{-4.1cm}
\caption{ The same as in Fig. 4 but now for initial (at $t =0$) $R_{\lambda} = 359$ instead of the initial $Re_{\lambda} = 143$.} 
\end{figure}
\begin{figure} \vspace{-0.4cm}\centering \hspace{-1cm}
\epsfig{width=.45\textwidth,file=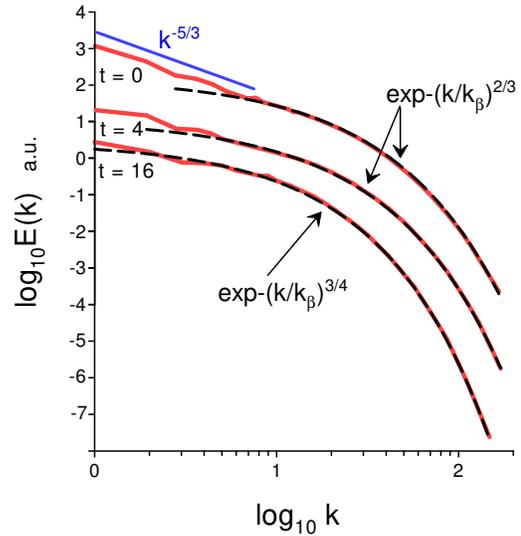} \vspace{-1.95cm}
\caption{ Kinetic energy spectra obtained in DNS of free decay starting from statistically stationary turbulence with $R_{\lambda} = 230$ (the spectra are shifted for clarity).} 
\end{figure}
  
   For the first Chkhetiani invariant itself (Eq. (11) for $p=0$) the estimate Eq. (18) should be replaced by 
$$
 v_c \propto  |^0\mathcal{I}^{\pm}|^{1/2}~ k_c,   \eqno{(27)}
$$  
i.e.  $\alpha =1$.  Then from the equation
 $$
 \beta = \frac{2\alpha}{1+2\alpha}   \eqno{(28)}
 $$
 (cf. Eq. (23)) one obtains $\beta = 2/3$, and
 $$
 E(k) \propto \exp(-k/k_{\beta})^{2/3}    \eqno{(29)}
 $$
 For a flow dominated by the Chkhetiani invariant $^1\mathcal{I}^{\pm} $ the estimation (27) should be replaced by
 $$
 v_c \propto  |^1\mathcal{I}^{\pm}|^{1/2}~ k_c^{3/2},   \eqno{(30)}
$$  
i.e. $\alpha = 3/2$. Then it follows from Eq. (28) that $\beta =3/4$ and
$$
 E(k) \propto \exp(-k/k_{\beta})^{3/4}.    \eqno{(31)}
 $$

      Since the Birkhoff-Saffman invariant Eq. (7) has the same dimensionality as the invariant $^1\mathcal{I}^{\pm} $ the same spectrum Eq. (31) can be obtained for a flow dominated by the Birkhoff-Saffman invariant. Whereas  for a flow dominated by the Loitsyanskii invariant Eq. (8) the analogous consideration results in $\beta =5/6$ and 
 $$
 E(k) \propto \exp(-k/k_{\beta})^{5/6}.    \eqno{(32)}
 $$     

   In Figs. 3-6 we have already seen a confirmation of the above-discussed variability of the distributed chaos. \\
   
   We can also see that the free decay generally results in an increase of the spectral parameter $\beta$ up to the value $\beta =1$ (which corresponds to deterministic chaos at small enough $R_{\lambda}$). Though, there can be a phenomenon of re-randomization when unstable deterministic chaos (with $\beta =1$) will appear prematurely. In this case further decay results in the re-appearance of distributed chaos which eventually will decay into deterministic chaos.\\

\begin{figure} \vspace{-1.03cm}\centering
\epsfig{width=.5\textwidth,file=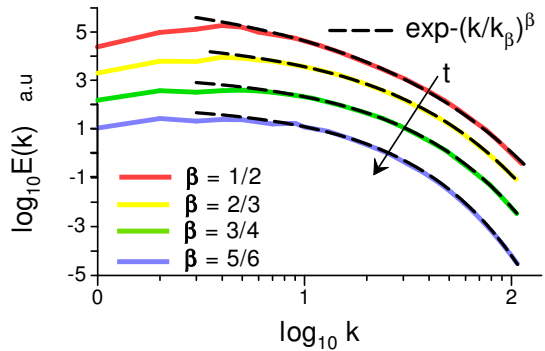} \vspace{-6.27cm}
\caption{ Kinetic energy spectra obtained in a DNS of free decay in a compressible fluid (the spectra are shifted for clarity).} 
\end{figure}

\begin{figure} \vspace{-0.5cm}\centering
\epsfig{width=.47\textwidth,file=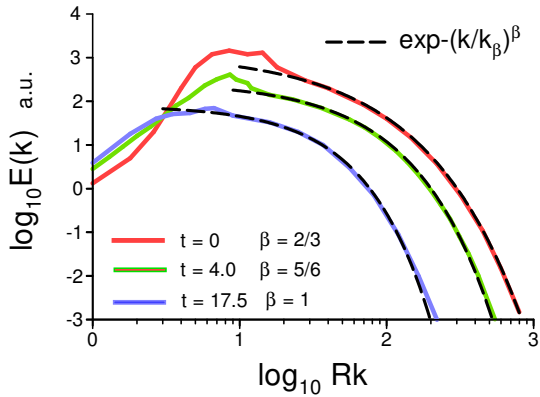} \vspace{-4.72cm}
\caption{ Kinetic energy spectra obtained in a DNS with an initially spherical region of chaos/turbulence freely evolving in space (the spectra are shifted for clarity).} 
\end{figure}
 
   Let us also consider several other examples of free decay variability. Figure 9 shows a picture of the energy spectrum decay for the same DNS as in Fig. 4 but now for initial $Re_{\lambda} = 359$ (instead of the initial $Re_{\lambda} = 143$ used for the DNS corresponding to Fig. 4). The spectral data were taken from Fig. 14 of the Ref. \cite{ym}. The $k_p$ corresponds to the peak value of the initial energy spectrum. The dashed curves indicate the exponential spectrum (deterministic chaos) for the `small' $Re_{\lambda} =41$, and stretched exponential spectra (distributed chaos) for moderate $Re_{\lambda}=78$ and $Re_{\lambda}=153$. One can see that again the criterion of `smallness' of $Re_{\lambda}$ can depend on the initial conditions. \\
   
    Results of a DNS of free decay starting from statistically stationary turbulence (after sudden turning off external forcing at the effective time $t=0$) were reported in a recent paper Ref. \cite{ypx}. \\

     Figure 10 shows kinetic energy spectra obtained at this DNS for $t = 0$ ($Re_{\lambda} =230$), for $t = 4$, and for $t = 16$ ($Re_{\lambda} =80$).  The spectral data were taken from Fig. 2a of Ref. \cite{ypx}. At $t = 0$ an indication of the appearance of a scaling (Kolmogorov) range of scales can be seen. This range corresponds to non-smooth dynamics. The next range of scales (with larger $k$) in this spectrum is characterized by the stretched exponential Eq. (29), i.e. it is dominated by the distributed chaos generated by the spontaneous breaking of local reflectional symmetry. \\
     
     At $t =4$ of the free decay the non-smooth dynamics has been smoothed and the scaling region disappears, whereas the same stretched exponential Eq. (29) is still present. Further free decay leads to the transformation of the spectrum Eq. (29) into the spectrum Eq. (31) (at $t = 6$, not shown in the Fig. 10). At the final time of the DNS $t = 16$ ($Re_{\lambda} = 80$) the spectrum Eq. (31) almost dominates the flow.\\
     
     It was already mentioned that the above consideration can be applied both to incompressible and compressible fluids. Figure 11 shows the evolution of the kinetic energy spectrum in free-decaying compressible fluid observed in a DNS \cite{spk}. The spectral data were taken from Fig. 3 of Ref. \cite{kps}. The DNS was performed in a periodic spatial cube with the initial spectrum taken as Eq. (3) with $n=4$. The Mach number $M \sim 0.4$,  and initial $Re_{\lambda} = 175$. The dashed curves indicate the stretched exponential spectra. One can see a picture typical also for incompressible fluids. \\

   An interesting DNS was reported in a recent paper Ref. \cite{yu}. This DNS was performed for an initially spherical region of chaos/turbulence freely evolving in space. The initial conditions comprise spherically windowed (inside a sphere of radius $R$), incompressible isotropic homogeneous chaos/turbulence. \\

   Figure 12 shows kinetic energy spectra obtained at this DNS for $t = 0$ ($Re_{\lambda} = 122.4$), $t = 4.0$, and $t = 17.5$. The spectral data were taken from Fig. 6 of the Ref. \cite{yu}. The dashed curves indicate the stretched exponential spectra for $t = 0$  Eq.(29), and for $t =4.0$ Eq. (32) (distributed chaos), and exponential spectrum (deterministic chaos) for the final time $t =17.5$ of the DNS. 
   
 \section{Free decay in magnetohydrodynamics}

    The equations for incompressible freely decaying magnetohydrodynamics in the Alfv\'enic units are
$$
 \frac{\partial {\bf u}}{\partial t} = - {\bf u} \cdot \nabla {\bf u} 
    -\frac{1}{\rho} \nabla {\cal P} - [{\bf b} \times (\nabla \times {\bf b})] + \nu \nabla^2  {\bf u} \eqno{(33)},
$$
$$
\frac{\partial {\bf b}}{\partial t} = \nabla \times ( {\bf u} \times
    {\bf b}) +\eta \nabla^2 {\bf b}  \eqno{(34)},
$$
where  ${\bf b} = {\bf B}/\sqrt{\mu_0\rho}$  is normalized magnetic field having the same dimensionality as velocity field ${\bf u}$ (also $  \nabla \cdot {\bf b} = 0$). \\

\begin{figure} \vspace{-0.65cm}\centering \hspace{-0.5cm}
\epsfig{width=.50\textwidth,file=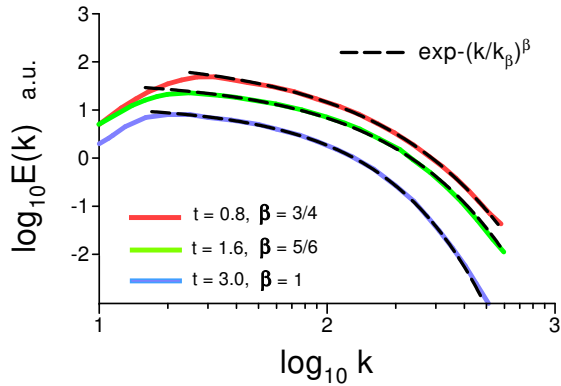} \vspace{-6.14cm}
\caption{ Kinetic energy spectra obtained in a magnetohydrodynamic DNS without external magnetic field (the spectra are shifted for clarity).} 
\end{figure}
\begin{figure} \vspace{-0.5cm}\centering \hspace{-1cm}
\epsfig{width=.45\textwidth,file=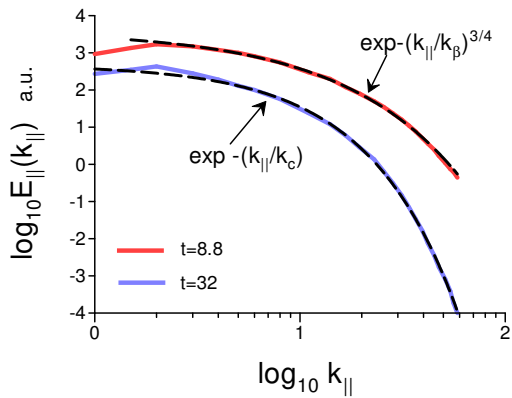} \vspace{-4.45cm}
\caption{ Kinetic energy spectra obtained in a magnetohydrodynamic large-eddy simulation with a uniform external magnetic field (the spectra are shifted for clarity).} 
\end{figure}
\begin{figure} \vspace{-0.9cm}\centering \hspace{-1cm}
\epsfig{width=.45\textwidth,file=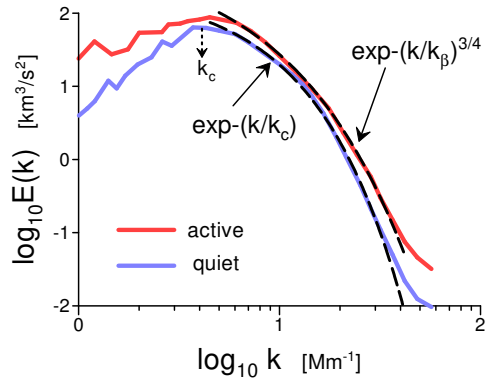} \vspace{-4.34cm}
\caption{ Spatial power spectra of the Doppler velocity for active and quiet solar photosphere regions.} 
\end{figure}

  The notion of distributed chaos was already applied to magnetohydrodynamics (see, for instance, papers Refs. \cite{b1},\cite{b2}). It was also shown in the papers Refs. \cite{b1}, \cite{cha} that the Loitsyanskii and Birkhoff-Saffman integrals are conserved in the freely decaying magnetohydrodynamics (a certain form of the Birkhoff-Saffman integral is conserved even in the presence of an external uniform magnetic field \cite{dav}). Therefore one can expect the spectra Eqs. (31-32) in the freely decaying magnetohydrodynamics as well. \\

   Results of a magnetohydrodynamic DNS with initial/boundary conditions similar to those used in Section II were reported in a recent paper Ref. \cite{abf}. Initial conditions for the velocity and magnetic fields were chosen to be the same (with $n = 4$ in the Eq. (3)), the equipartition case. \\
   
   Figure 13 shows the kinetic energy spectra for a so-called nonhelical case (the data were taken from Figs. 1 and 14 of the Ref. \cite{abf}).  The initial Reynolds number $Re = 129$ and the magnetic Prandtl number $Pm = \nu/\eta =1$ in the considered case. The dashed curves indicate the stretched exponential spectra for $t = 0.8$  Eq. (31), for $t =1.6$ Eq. (32) (distributed chaos), and exponential spectrum (deterministic chaos) for $t =3$. \\

    Figure 14 shows the kinetic energy spectra obtained in the asymptotic limit of low magnetic Reynolds numbers using a large-eddy simulation (the dynamic Smagorinsky model) and reported in the paper Ref. \cite{bzk}. The spectral data were taken from Fig. 33 of the Ref. \cite{bzk} (initial $Re_{\lambda} = 170$).  The index $||$ denotes the components parallel to the external uniform magnetic field ${\bf B}_0$. The dashed curves indicate the stretched exponential spectrum for $t =8.8$  Eq. (31) (distributed chaos) and exponential spectrum Eq. (1)  (deterministic chaos) for $t =32$.\\

  Figure 15 shows the power spectra of the Doppler velocity measured in active and quiet regions of the solar photosphere. The spectral data were taken from Fig. 3 of Ref. \cite{cha}. The measurements were produced with the balloon-borne Sunrise missions. The quiet region was located at the solar disk center and the trailing part of active region AR11768 was observed at heliocentric angle $\mu = 0.93$. The solar photosphere is highly stratified (see, for instance, Ref. \cite{ss}). However, similar to the case of an external uniform magnetic field, in the cases of stratification and rotation a generalization of the Birkhoff-Saffman invariant is still valid \cite{dav}. The dashed curves indicate correspondence to the Eq. (1) (deterministic chaos in the quiet solar region) and to the Eq. (31) (the Birkhoff-Saffman distributed chaos in the active solar region), cf Fig. 14. Location of the $k_c$ at the peak of the spectrum (the dotted arrow) indicates that the large-scale coherent structures determine the deterministic chaos in the quiet solar region.

\end{document}